\documentclass[aps,prx,twocolumn,superscriptaddress]{revtex4-2}

\usepackage{amsmath,amssymb,amsfonts,amsthm}
\usepackage{graphicx,subfigure}% Include figure files
\usepackage{dcolumn}% Align table columns on decimal point
\usepackage{bm,bbm}% bold math
\usepackage{hyperref}% add hypertext capabilities
\usepackage{color}
\usepackage[section]{placeins}
\usepackage{setspace}
\usepackage{enumitem}

\makeatletter
\renewcommand\frontmatter@abstractwidth{\dimexpr\textwidth-0.25in\relax}
\makeatother

\usepackage[T1]{fontenc}

\begin{document}
	
% number in the upper right hand corner of the title page in preprint mode.
%\preprint{}
	
\title{Realization of all-optical underdamped stochastic Stirling engine}


	
\author{Chuang Li}
\thanks{These authors contributed equally to this work.}
\affiliation{Research Center for Quantum Sensing, Intelligent Perception Research Institute, Zhejiang Lab, Hangzhou, 311121, China}

\author{Shaochong Zhu}
\thanks{These authors contributed equally to this work.}
\affiliation{Research Center for Quantum Sensing, Intelligent Perception Research Institute, Zhejiang Lab, Hangzhou, 311121, China}

\author{Peitong He}
\affiliation{Research Center for Quantum Sensing, Intelligent Perception Research Institute, Zhejiang Lab, Hangzhou, 311121, China}

\author{Yingying Wang}
\affiliation{Research Center for Quantum Sensing, Intelligent Perception Research Institute, Zhejiang Lab, Hangzhou, 311121, China}

\author{Yi Zheng}
\affiliation{Research Center for Quantum Sensing, Intelligent Perception Research Institute, Zhejiang Lab, Hangzhou, 311121, China}

\author{Kexin Zhang}
\affiliation{Research Center for Quantum Sensing, Intelligent Perception Research Institute, Zhejiang Lab, Hangzhou, 311121, China}

\author{Xiaowen Gao}
\email[]{gaoxw@zhejianglab.com}
\affiliation{Research Center for Quantum Sensing, Intelligent Perception Research Institute, Zhejiang Lab, Hangzhou, 311121, China}

\author{Ying Dong}
\email[]{yingdong@zhejianglab.com}
\affiliation{Research Center for Quantum Sensing, Intelligent Perception Research Institute, Zhejiang Lab, Hangzhou, 311121, China}

\author{Huizhu Hu}
\affiliation{Research Center for Quantum Sensing, Intelligent Perception Research Institute, Zhejiang Lab, Hangzhou, 311121, China}
\affiliation{State Key Laboratory of Modern Optical Instrumentation $\mathrm{\&}$ College of Optical Science and Engineering, Zhejiang University, Hangzhou, 310027, China}

	
\date{\today}
	
\begin{abstract}
We experimentally demonstrate a nano-scale stochastic Stirling heat engine operating in the underdamped regime. The setup involves an optically levitated silica particle that is subjected to a power-varying optical trap and periodically coupled to a cold/hot reservoir via switching on/off active feedback cooling. We conduct a systematic investigation of the engine's performance and find that both the output work and efficiency approach their theoretical limits under quasi-static conditions. Furthermore, we examine the dependence of the output work fluctuation on the cycle time and temperature difference between the hot and cold reservoirs. We observe that the distribution has a Gaussian profile in the quasi-static regime,  whereas it becomes asymmetric and non-Gaussian as the cycle duration time decreases. This non-Gaussianity is qualitatively attributed to the strong correlation of the particle’s position within a cycle in the non-equilibrium regime. Our experiments provide valuable insights into stochastic thermodynamics in the underdamped regime and open up new possibilities for the design of future nano-machines.

%Our numerical study verifies that the non-Gaussian work distribution causes by the position correlation of different instants, which is enhanced for a short cycle time.     
%also dependent on the temperature difference between the hot and cold reservoirs.
% An obvious fluctuation of output work in individual cycles is observed due to the stochastic nature of the Brownian motion of the levitated particle. Contrary to intuition, the output work is not simply a Gaussian distribution centered at the average value, as observed previously from an overdamped stochastic heat engine. The deviation from Gaussian distribution is investigated systematically with numerical simulations, which confirm that the non-Gaussian distributed output work is a special feature of a stochastic heat engine in the underdamped regime, resulting from the immediate response of the oscillating amplitude to the varying of the trapping stiffness.
\end{abstract}
	
%\keywords{}

\maketitle

Thermodynamics deals with the relations between heat, work, temperature, entropy, and energy \cite{Fermi}. At its heart, is the heat engine, operating periodically between two reservoirs with different temperatures. Unlike its macroscopic counterpart that the deterministic classical thermodynamic laws can very well describe, a heat engine of micro- or nano-size will undergo visible fluctuations \cite{Spinney2012}, which makes it behave in a stochastic manner. In this regime, central concepts of thermodynamics such as the exchanged heat, the applied work, and the entropy can be meaningfully defined on the level of individual trajectories  \cite{Seifert2005,Seifert2012}. These fluctuating quantities extend the laws of macroscopic thermodynamics and give birth to the so-called stochastic energetics \cite{Sekimoto2010}. With the advancement in the fabrication of microscopic mechanical devices, significant progress in the study of stochastic heat engines, both theoretical \cite{Dechant2015,Dong2015} and experimental \cite{Ciliberto2017,Wu2021}, have been witnessed during the past decade.

A promising candidate for the experimental investigation of the stochastic thermodynamics \cite{Gieseler2013,Gieseler2014,Rondin2015,Gieseler2018,Rademacher2022} is the levitated optomechanical system (LOS) \cite{Millen2020,Romero-Isart2021}, where optical tweezers allow us to apply a fast and accurate control to the particle captured and record its spatial trajectory in real-time \cite{Ashkin1976,Ashkin1986}.
After a full description of a colloidal stochastic heat engine \cite{Seifert2008} given by Schmiedl and Siefert, Blickle and Bechinger realized a stochastic Stirling engine experimentally \cite{Blickle2011} for the first time. The efficiency of the Stirling engine is fundamentally limited by the isochoric steps, which make the cycle inherently irreversible. To overcome the limitations of the Stirling cycle, Martinez et al. implemented a Brownian Carnot cycle \cite{Martinez2015} with an optically trapped colloidal particle by creating an effective hot temperature bath with fluctuating electromagnetic fields, which allowed precise control over the bath temperature that is synchronized with the change of the trap stiffness and therefore a realization of an adiabatic ramp.

So far, the implementations of stochastic heat engines (SHEs) with a levitated microscopic particle are all in colloidal systems where the particle is overdamped \cite{imparato2007work,Martinez2017}. The SHEs in the underdamped regime have been less investigated experimentally yet. We know that an analytic treatment of optimal protocols is possible in the overdamped case because the dynamics can be described by a simplified equation in terms of the slow position variable \cite{Seifert2008}. In contrast, this is not possible in the underdamped case, where the position and velocity variables cannot be separated. As a result, the optimization of an underdamped SHE for maximizing its performance is much more complicated than an overdamped one. Theoretical analysis showed that rapid changes in the trapping frequency were desired to improve the power output and the efficiency \cite{Lutz2017}. More importantly, the investigations of much more isolated systems provide a path toward the future realization of quantum heat engines \cite{Zhang2014,Zhang2015} or quantum refrigerators \cite{Dong2015prl} in LOS and it has been shown that super Carnot efficiencies can be attained by clever reservoir engineering \cite{Scully2010,prx,Niedenzu2018}. 

\begin{figure}[htbp]
	\centering
	\includegraphics[width=0.45\textwidth]{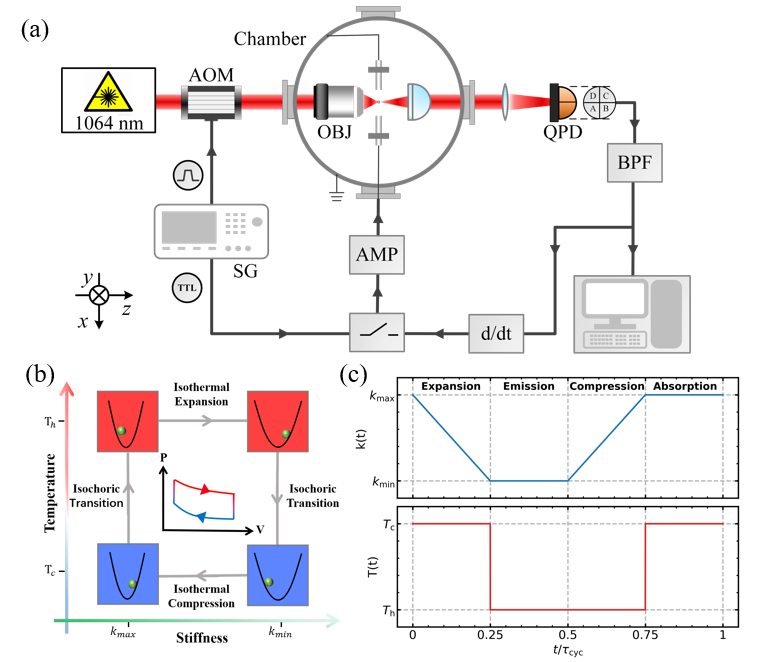}
        \caption{(a) The experimental setup: A laser beam of wavelength $1064$ nm passes an acousto-optic modulator (AOM) and is focused by a microscope objective (OBJ) with numerical aperture $\mathrm{NA} = 0.8$ forming an optical trap, which levitates a charged silica particle inside a vacuum chamber. The optical trap stiffness (or frequency) is dependent on the optical trap power which can be changed by adjusting the driving voltage of the AOM. Throughout our work, we solely focus on the motion of the particle along the $x$ axis. The scattered light from the particle is collected and sent to a quadrant photodetector (QPD) to detect the motion of the particle along the $x$ axis. A feedback cooling scheme based on electric fields is applied to create the cold bath. In the scheme, the $x$-motion signal is sent through a bandpass filter (BPF) and a derivative circuit ($\mathrm{d}/\mathrm{d}t$) to provide a feedback signal proportional to velocity. This velocity-dependent feedback signal is sent to an amplifier (AMP) which modulates a pair of electrodes to cool the x-motion electrically. To experimentally realize a stochastic Stirling cycle, we employ a signal generator (SG) that periodically outputs two synchronized signals. One signal is sent to the AOM to linearly change the optical trap power, and the other one is sent to a switch to periodically turn on/off the feedback cooling. (b) The Stirling cycle consists of two isochoric and two isothermal strokes. The inset shows the analogy to a classical Stirling engine. (c) The optical trap stiffness $k(t)$ and the temperature of COM motion $T(t)$ as a function of time $t$ during a Stirling cycle.}
	\label{fig:experiment_setup}
\end{figure}

Inspired by a scheme \cite{Dechant2015} proposed by Dechant et al., we present an experimental realization of an all-optical Stirling engine in the underdamped regime. The experimental setup is illustrated in Fig. \ref{fig:experiment_setup}(a).
A charged silica particle of diameter $153.6$ nm is levitated in a single-beam optical trap at the pressure of $p = 1.0$ mbar.
The damping rate due to the collision with residual gas molecules is experimentally measured as $\Gamma_{\mathrm{th}}/2\pi = 1.45$ kHz.
%frequency $\omega_{\mathrm{m}}$ from $140.1$ kHz to $160.3$ kHz
The optical potential is approximately harmonic with a power-dependent stiffness. We thus use an acousto-optic modulator (AOM) to linearly change the optical trap power resulting in a linear variation of the optical trap stiffness. The resulting mechanical oscillation frequency of the particle $\Omega/2\pi$ ranges from $145.3$ to $160.4$ kHz.
The surrounding gas environment coupling to the center of mass (COM) motion of the particle acts as the high-temperature reservoir (hot bath).
The active feedback cooling, using electric fields to exert a force on the particle’s motion and providing an additional feedback damping rate $\Gamma_{\mathrm{fb}}$, creates a controllable low-temperature reservoir (cold bath).
A quadrant photodetector (QPD) monitors the motion of the particle so that its position trajectory can be accurately recorded.
To realize a stochastic Stirling cycle consisting of two isothermal and two isochoric strokes, we employ a signal generator (SG) to periodically output two synchronized signals, which are inputted into the AOM to change the optical trap power and into the switch to turn on/off the feedback cooling, respectively.
The detail of the experimental setup and protocol can be found in the caption of Fig. \ref{fig:experiment_setup} and the Supplemental Material (SM).

%In our feedback cooling scheme, the coulomb force acts as the feedback force which provides an additional tunable feedback damping  and can control the temperature of COM motion.  
% In combination with the feedback cooling scheme, the temperature of  can be controlled. Here, the coulomb force acts as the feedback force providing     
%by using a silica nanoparticle optically trapped in high vacuum. The experimental scheme is shown in Fig. \ref{fig:experiment_setup}(a). In this system, 
% the frequency of the harmonic potential is determined by the laser intensity which can be fast controlled with an acousto-optic modulator (AOM). The heat baths are created by a thermal environment (from the surrounding gas) in combination with parametric feedback cooling \cite{Gieseler2012}, which provides an additional tunable damping rate of the particle. The one-dimension position trajectory of the particle can be accurately recorded through balanced detection \cite{Li2010}. 
% These distinct advantages of the LOS under high vacuum conditions make it very suited to investigate the SHEs in the underdamped regime.
% \textit{The model.} We consider an optically levitated nanoparticle of mass $m$ confined in a one-dimensional harmonic trap potential with the trapping frequency $\Omega$. The parametric feedback cooling is applied to the nanosphere and provides a controllable additional damping such that its motion of center-of-mass (COM) can be effectively coupled to the reservoirs with different temperatures.
\textit{The model.}---As described above, the particle's oscillation frequency is much larger than the damping rate. Therefore, its COM motion is governed by the one-dimensional underdamped Langevin equation \cite{aurell2017work} (we solely focus on the particle's motion along the $x$ axis throughout this work),
% \begin{equation}
%     m\dot{v} + m\gamma v + k(t)x^2 = \sqrt{2 m k_{B} T \Gamma}\xi(t), 
% \end{equation}
\begin{equation}
	\dot{v} + \Gamma v + k(t) x = \sqrt{2 k_{B} T \Gamma /m } \xi(t),
\end{equation}
%distance deviating from the trapping center
where $x$ and $v = \dot{x}$ respectively denote the position and velocity of the particle. $k(t) = m \Omega^2$ denotes the optical trap stiffness with the particle's mass $m$ and frequency $\Omega$. $\Gamma = \Gamma_{\mathrm{th}} +\Gamma_{\mathrm{fb}}$ is the total damping rate, $T$ is the effective COM temperature, and $k_{\mathrm{B}}$ is the Boltzmann constant. The quantity $\xi(t)$ is a centered Gaussian white noise with $\langle \xi(t)\xi(t^\prime)\rangle = \delta(t - t^\prime)$.
%The specific steps for a stochastic Stirling cycle are as follows:
%To realize a stochastic Stirling cycle consisting of two isothermal and two isochoric strokes, we employ a signal generator (SG) to periodically output two synchronized signals, which are inputted into the AOM to change the optical trap power and into the switch to turn on/off the feedback cooling, respectively (see SM).
% originated from the gas-nanosphere collisions $\Gamma_{\mathrm{th}}$ together with the feedback cooling  $\Gamma_{\mathrm{fd}}$ (if applied).  
%\noindent\textbf{The Stirling cycle.}
% By varying the power of the optical trap, the stiffness of the optical spring defined as $k(t) = m\Omega^2$ can be tuned periodically. 
A Stirling cycle consisting of two isochoric and two isothermal strokes will be realized in the following steps:
% \begin{enumerate}
\begin{itemize}[leftmargin=*]
\item Expansion. Starting from the system in thermal equilibrium with the hot bath at the temperature $T_{\mathrm{h}}$, it undergoes an isothermal expansion with stiffness $k(t)$ decreasing from $k_{\mathrm{max}}$ to $k_{\mathrm{min}}$ during a time period $\tau_{\mathrm{h}}$. The particle keeps the connection to the hot reservoir via the coupling $\Gamma_{\mathrm{th}}$ during this stroke. To make sure that the stroke is isothermal, we require that the duration $\tau_{\mathrm{h}} \gg 1/\Gamma_{\mathrm{th}}$ and the oscillator therefore always equilibrates with the reservoir.

\item Heat emission. The temperature of the reservoir is reduced to $T_{\mathrm{c}}  = \Gamma_{\mathrm{th}}T_{\mathrm{h}}/(\Gamma_{\mathrm{th}} + \Gamma_{\mathrm{fd}})$ by switching on the feedback cooling. In this stroke, the particle is instantaneously connected to the cold bath while the stiffness retains constant $k(t) = k_{\mathrm{min}}$ lasting a time duration of $\tau_{\mathrm{hc}}$ until its effective temperature of COM equilibrates to $T_{\mathrm{c}}$.

\item Compression. The system undergoes an isothermal compression with the stiffness $k(t)$ increasing back to $k_{\mathrm{max}}$ during a time duration $\tau_{\mathrm{c}}$. In this stage, the oscillator always keeps the connection to the cold bath at the constant temperature $T_{\mathrm{c}}$.

\item Heat absorption. The feedback cooling is switched off so that the particle is connected to the hot reservoir again and reaches back to the equilibrium state at the beginning of the cycle in a period of $\tau_{\mathrm{ch}}$. The stiffness keeps constant $k(t) = k_{\mathrm{max}}$ during this stroke.
\end{itemize}
% \end{enumerate}
	
The total duration period of a cycle is then given by $\tau_{\mathrm{cyc}} = \tau_{\mathrm{h}} + \tau_{\mathrm{hc}} +  \tau_{\mathrm{c}} + \tau_{\mathrm{ch}}$. Here, we set the four strokes with the equal duration $\tau_{\mathrm{s}}$, i.e., the cycle time $\tau_{\mathrm{cyc}} = 4\tau_{\mathrm{s}}$.
The schematic representation for a cyclic process of the Stirling engine and 
the change of the trap stiffness and the effective COM temperature versus time is shown in Fig. \ref{fig:experiment_setup} (b) and (c), respectively.
%The schematic representation for a cyclic process of the Stirling engine is shown in Fig. \ref{fig:experiment_setup}(b).
% where the time duration of temperature change $\tau_{\mathrm{hc}}$ and $\tau_{\mathrm{ch}}$ can actually be very short due to fast feedback cooling.

One can draw an analogy between the particle in an optical trap and an ideal gas inside a piston, where the trap stiffness, or equivalently the trapping frequency, is analogous to the inverse of an effective volume while the variance of the particle position is seen as an effective pressure. Under this analogy, thermodynamic quantities can be extracted from the particle's positional fluctuations in the framework of stochastic thermodynamics. The total energy of the particle at time $t$ reads
\begin{equation}
	U(t) = \frac{1}{2} m v(t)^2 + \frac{1}{2}  k(t) x(t)^2.
\end{equation}
The increment of the energy $\mathrm{d}U$ can thus be divided into two parts where the output work is defined as
\begin{equation}\label{dwork}
	\mathrm{d}W = \frac{1}{2}\frac{\mathrm{d}k}{\mathrm{d}t} x^2 \mathrm{d}t
\end{equation}
and the heat exchanged with the environment is defined as
\begin{equation}
	\mathrm{d}Q = m v \frac{\mathrm{d}v}{\mathrm{d}t}\mathrm{d}t + kx\frac{\mathrm{d}x}{\mathrm{d}t}\mathrm{d}t.
\end{equation}
Integrating Eq. (\ref{dwork}) along a stochastic trajectory yields the time-dependent work during a time duration $\tau = t_{\mathrm{f}} - t_{\mathrm{i}}$ as
\begin{equation}\label{eq: work}
	W(\tau) = \frac{1}{2}\int_{t_{\mathrm{i}}}^{t_{\mathrm{f}}}\frac{\mathrm{d} k(t)}{\mathrm{d} t} x(t)^2 \mathrm{d}t,
\end{equation}
where $t_{\mathrm{i}}$ ($t_{\mathrm{f}}$) is the initial (final) time.
Meanwhile, the work $W$, heat $Q$, and inner energy $U$ satisfy the stochastic first-law-like energy balance $\Delta U = W + Q$ for any single stochastic trajectory. 
%One thus can obtain the exchanged heat $Q$ during the time duration $\tau$ via
%\begin{equation}
%	Q(\tau) = U(t_{\mathrm{f}}) - U(t_{\mathrm{i}}) - W(\tau).
%\end{equation}
%The equation (\ref{eq: work}) is in agreement with a macroscopic Stirling engine.
%From Eq. (\ref{eq: work}), one can find that the two isochoric strokes in the Stirling cycle do not contribute to the work $W$ due to the constant stiffness $k(t)$.
%where $t_{\mathrm{i}}$ ($t_{\mathrm{f}}$) is the initial (final) time and $w(\tau)$ is the work during the duration $\tau$ which can be calculated by Eq. (\ref{eq: work}).

\begin{figure}[htbp]
	\centering
	\includegraphics[width=0.49\textwidth]{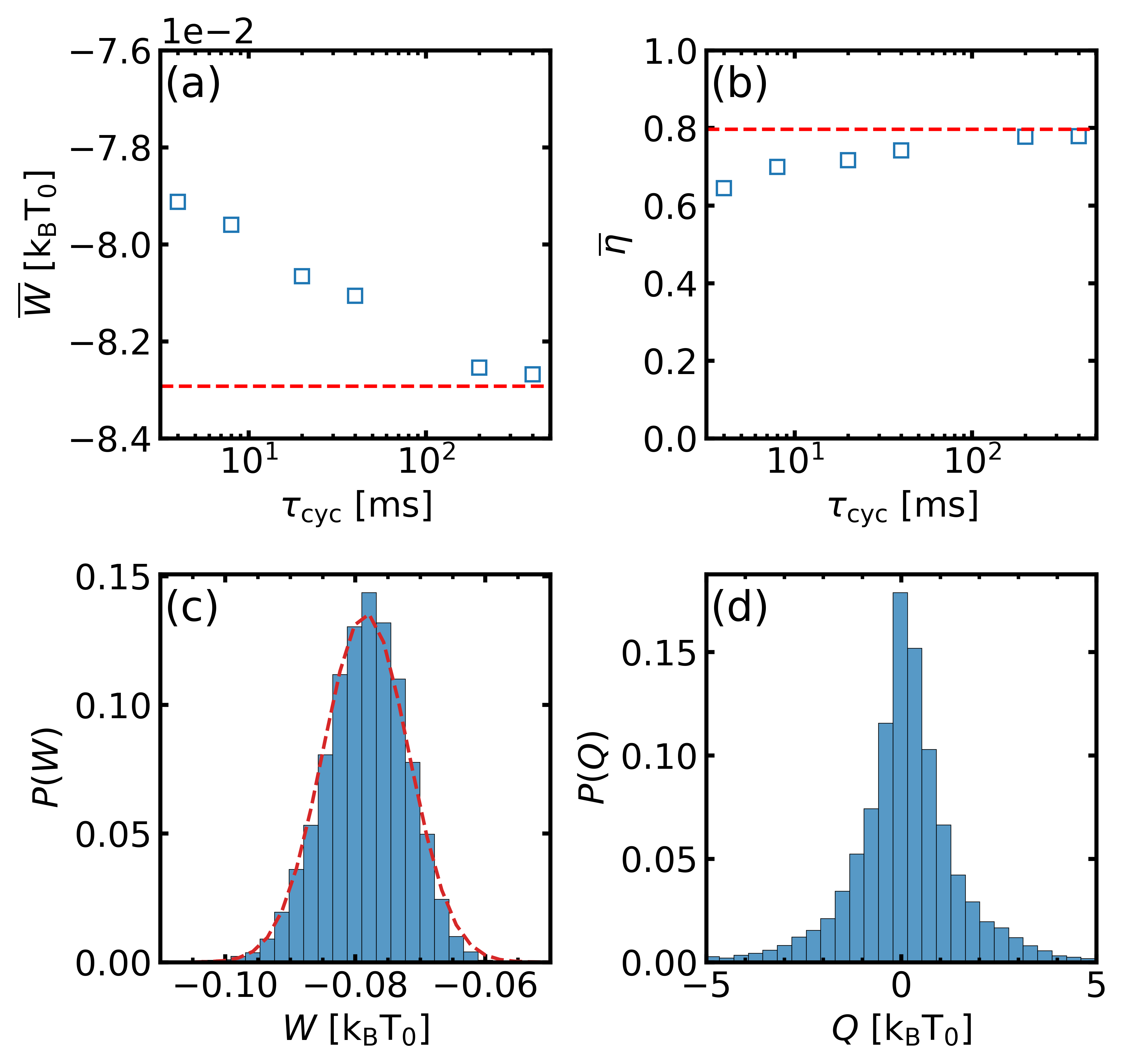}
	\caption{(Color online) (a) The mean output work $\overline{W}$ and (b) the mean efficiency $\overline{\eta} = -\frac{\overline{W}}{\overline{Q}}$ as a function of the cycle time $\tau_{\mathrm{cyc}}$. Here, the cycle time of the experiment is chosen as $\tau_{\mathrm{cyc}} = 4,\ 8,\ 20,\ 40,\ 200,\ 400$ ms. The red dashed lines in (a-b) are respectively the theoretical values of the output work and efficiency in the quasi-static regime (see the SM for detail).  The probability distributions of (c) the output work $W$ and (d) the absorbed heat $Q$ in the quasi-static regime for the cycle time $\tau_{\mathrm{cyc}} = 400$ ms. The red dashed line in (c) is a Gaussian distribution with the same mean and variance of the output work distribution.
	Throughout this paper, all the work and the heat are in units of $k_{\mathrm{B}} T_0$.
	The temperature of the particle's COM motion touching with the hot and cold bath is experimentally determined as $T_h = 320$ K and $T_c = 65$ K, respectively.
	The number of performing Stirling cycles (ensemble number) in each experiment is fixed as $N = 10000$.}
	\label{fig:cycle time}
\end{figure}
%\newline
%{\fontfamily{phv}\selectfont \noindent\textbf{Results}}

\textit{The experiments.}---We experimentally studied the performance of the Stirling engine in the underdamped regime. The optical stiffness linearly change in the range from $k_{\mathrm{min}} = 3.29$ aN/$\mathrm{\mu}$m ($\sim \Omega_{\mathrm{min}}/2\pi = 145.3$ kHz) to $k_{\mathrm{max}} = 3.63$ aN/$\mathrm{\mu}$m ($\sim \Omega_{\mathrm{max}}/2\pi = 160.4$ kHz).
The effective COM temperature of the levitated particle is experimentally determined as $T_h = 320$ K and $T_c = 65$ K, respectively. Here, the temperature equilibrated to the hot bath is a little bit larger than the room temperature $T_0 = 300$ K due to the heating effect of the trapping laser \cite{millen2014nanoscale}.
%The shortest stroke time subject to our experimental setup is $\tau_{\mathrm{s}} = 0.5$ ms which is smaller than the thermal relaxation time $\Gamma_{\mathrm{th}}^{-1} \approx 0.7$ ms, allowing us to explore the nonequilibrium thermodynamics.
% Our experimental setup limits the shortest cycle time $\tau_{\mathrm{s}} = 0.5$ ms which is smaller than the relaxation time $\gamma_{p}^{-1} \approx 0.7$ ms.
Figure \ref{fig:cycle time}(a-b) show the mean output work $\overline{W}$ and the mean efficiency $\eta = -\frac{\overline{W}}{\overline{Q}}$ varying with the cycle time $\tau_{\mathrm{cyc}}$, respectively.
%the mean power $\overline{P} = -\frac{\overline{W}}{\tau_{\mathrm{cyc}}}$,
The work and heat are both in units of $k_{\mathrm{B}}T_0$ throughout this paper.
One can see, both the mean output work $\overline{W}$ and the mean efficiency $\eta$ monotonically increase with the cycle time and finally converge to the theoretical limit $\overline{W}_{\mathrm{qs}} = -\frac{1}{2}\frac{T_{\mathrm{h}} - T_{\mathrm{c}}}{T_0} \ln \frac{k_{\mathrm{max}}}{k_{\mathrm{min}}}$ \cite{rana2014single} and the Carnot efficiency $\eta_{\mathrm{qs}} = 1 - \frac{T_c}{T_h} = 0.79$ respectively in the quasi-static regime with an infinite long cycle time. By numerical simulation, we also figured out that the maximum power $\overline{P} = -\frac{\overline{W}}{\tau_{\mathrm{cyc}}}$ will reach its maximum value \cite{Blickle2011} at $\tau_{\mathrm{cyc}} \approx 0.04$ ms.  It is not observed since $0.04$ ms is much less than the shortest cycle time $\tau_{\mathrm{cyc}} = 4$ ms that is allowed in our experiments. The theoretical and simulation results and the detailed discussion of the power are presented in the SM.

Figure \ref{fig:cycle time} (c-d) show the probability distributions of the output work $W$ and the absorbed heat $Q$ in the quasi-static regime for $\tau_{\mathrm{cyc}} = 400$ ms. As expected, strong fluctuations can be seen in the output work and the absorbed heat. We compare the measured output work distribution with a Gaussian distribution with the same mean and variance in Fig. \ref{fig:cycle time}(c). We also calculated the Jensen-Shannon divergence (JSD) \cite{JSD1} between them as  $D_{\mathrm{JS}} = 0.0043$ (see the SM), which confirms that the output work distribution in the quasi-static regime is Gaussian.

\begin{figure}
	\centering
	\includegraphics[width=0.49\textwidth]{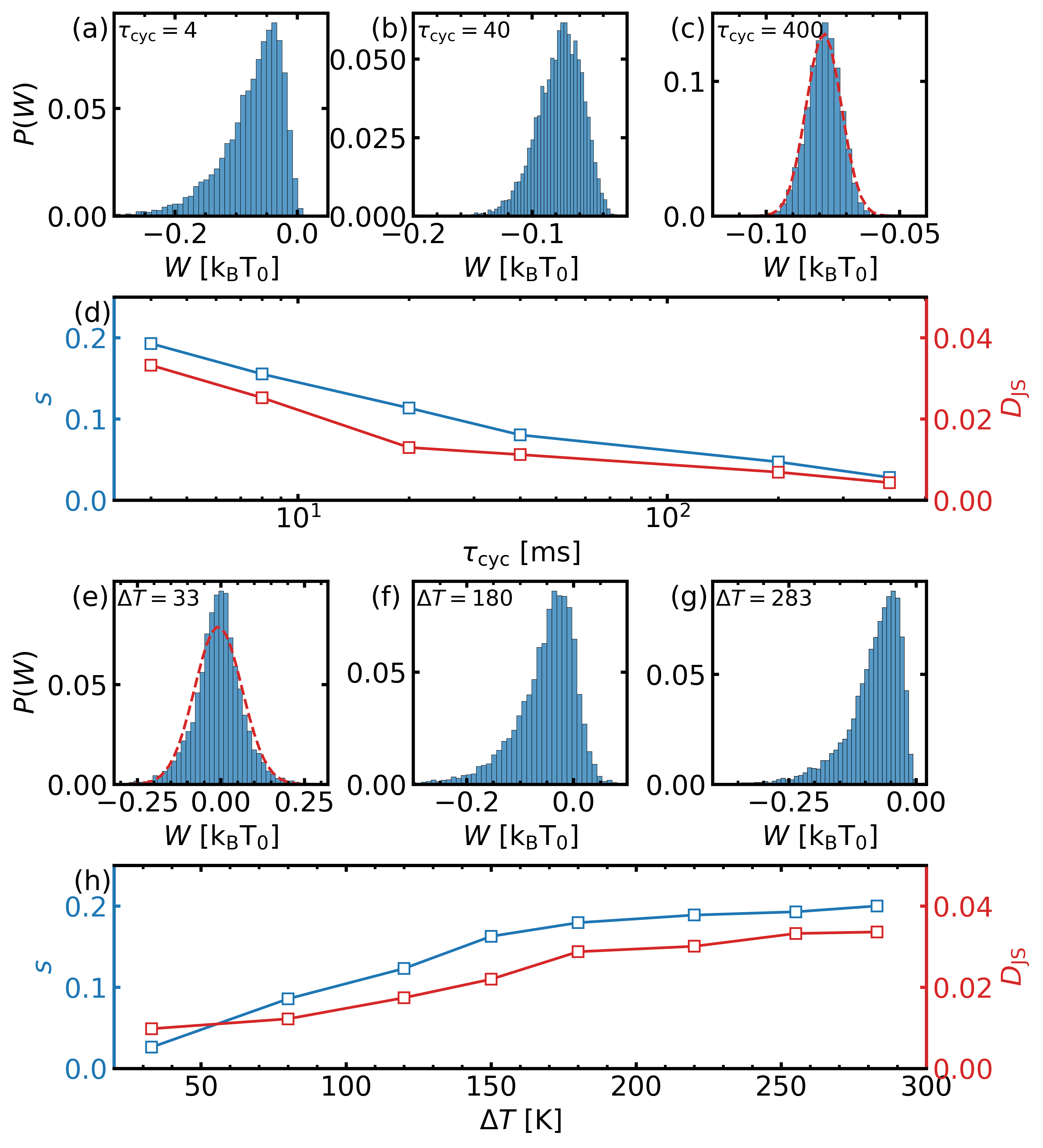}
	\caption{(Color online) (a-c) The probability distributions of the output work for the cycle time (a) $\tau_{\mathrm{cyc}} = 4$ ms, (b) $\tau_{\mathrm{cyc}} = 40$ ms, and (c) $\tau_{\mathrm{cyc}} = 400$ ms (in the quasi-static regime).
	(d) The symmetric index $s$ and JSD as a function of the cycle time $\tau_{\mathrm{cyc}}$. In (a-d), the temperature difference between the hot and cold baths is fixed as $\Delta T = T_{\mathrm{h}} - T_{\mathrm{c}}= 255$ K.
	(e-g) The probability distribution of the output work for the temperature difference (e) $\Delta T = 33$ K,  (f) $\Delta T = 180$ K, and (g) $\Delta T = 283$ K.
	(h) The symmetric index $s$ and JSD as a function of the temperature difference $\Delta T$.
	In (e-h), the cycle time is fixed as $\tau_{\mathrm{cyc}} = 4$ ms.  The red dashed lines in (c) and (e) are a Gaussian distribution with the same mean and variance of the output work distribution.}
	\label{fig:symmetry}
\end{figure}

We further explore the output work distribution for different cycle times, as shown in Figure \ref{fig:symmetry}(a-c). Interestingly, we found that the profile of the work distribution becomes asymmetric as we decrease the cycle time.
%and $\tau_{\mathrm{cyc}} = 400$ ms. Contrary to the distribution in the quasistatic regime, the distribution of $\tau_{\mathrm{cyc}} = 2$ is obviously asymmetric, that is, the symmetry of the work distribution is broken near (or in) the non-equilibrium regime.
% we observed the non-Gaussian-shape distribution of the output work that was not reported in the previous experimental studies of overdamped SHEs. The symmetry of the distribution was broken when the temperature difference is large.
%Intuitively, we may expect the output work distributed in a symmetric Gaussian form. 
%But in the experiments, we found that the symmetry was broken when the temperature difference between the hot and cold reservoirs is large.
To quantify the symmetry breaking of work distribution, we define a symmetric index $s$ simply as
\begin{equation}
	s = \frac{|N_{\mathrm{la}} - N_{\mathrm{le}}|}{N_{\mathrm{la}} + N_{\mathrm{le}}},
\end{equation}
where $N_{\mathrm{la}}$ ($N_{\mathrm{le}}$) is the count of cycles in which the value of the output work is larger (less) than the mean. The index $s$ ranges from $0$ (corresponding to a symmetric distribution) to $1$ (corresponding to a fully asymmetric distribution).
Fig. \ref{fig:symmetry}(d) shows the symmetric index $s$ and the JSD between the measured work distribution and the Gaussian distribution with the same mean and variance versus the cycle time. Both of them increase as the cycle time decreases, indicating that the output work will deviate from the Gaussian distribution \cite{kwon2013work} when the engine operates under non-quasi-static conditions.

%The inset in Fig. \ref{fig:symmetry}(a) shows the index $s$ versus the cycle time $\tau_{\mathrm{cyc}}$. As we decrease the cycle time away from the equilibrium, the work distribution progressively becomes asymmetric.

Moreover, we found that the work distribution profile also depends on the temperature difference $\Delta T = T_{\mathrm{h}} - T_{\mathrm{c}}$ between the hot and cold baths.
%near the non-equilibrium regime ($\tau_{\mathrm{cycl}} = 2$ ms). 
Figure \ref{fig:symmetry}(e-g) shows how the work distribution changes with varying the temperature difference $\Delta T$ for the short cycle time $\tau_{\mathrm{cyc}} = 4$ ms, and the index $s$ and JSD versus the temperature difference is plotted in (h).
The results show that the work distribution with a short cycle time can change from asymmetric to symmetric with decreasing temperature differences. However, the symmetric distribution is still non-Gaussian with a visible JSD. 
%of the large temperature difference $\Delta T = 287$ K and the small temperature difference $\Delta T = 36$ K, and the inset gives the index $s$ varying with the temperature difference $\Delta T$. From the results, we found that the distribution becomes symmetric again when the temperature difference is small in the non-equilibrium regime. The trend of symmetry index $s$ clearly shows that the distribution  progressively becomes symmetric as we decrease the temperature difference.  
% We also calculated the Jensen-Shannon divergence \cite{JSD1} between the obtained distribution and the standard Gaussian distribution with the same mean and variance to quantify the non-similarity between them (See Supplemental Materials). The results show that both the $s$ and JSD increase with the increase of the temperature difference (see (c)). 
%In addition, both the mean output work $\overline{w}$ and the mean efficiency $\overline{\eta}$ increase with increasing the temperature difference $\Delta T$ (see (d-e)). The mean efficiency $\overline{\eta}$ has reached $73\%$ in the case of $\Delta T = 286$ K.
% We also experimentally investigated the performance of the underdamped Stirling engine for different stiffness difference $\Delta k = k_{\mathrm{max}} - k_{\mathrm{min}}$ and cycle time $\tau_{\mathrm{cycle}}$ and the results are shown in Fig.
	
% \textit{The simulations.}
% To confirm the asymmetric work distribution is indeed a special feature of SHEs in the underdamped regime rather than a statistical error, we performed numerical simulations (see  Supplemental Materials for details) of SHEs in both underdamped and overdamped cases. 

% The dependence of $s$ as on the temperature $\Delta T$ in the underdamped and overdamped regimes are shown in Fig. \ref{fig:simulation_results}. The results clearly confirm that the asymmetry of work distribution is intensified with the increasing of the reservoir temperature difference in the underdamped regime, while the distribution keeps symmetric in the overdamped regime. Compared to the experimental results, we can actually see a much more obvious non-Gaussian distribution directly in the underdamped case when the temperature difference is large, as shown in the inset figures. 

\begin{figure}
	\centering
	\includegraphics[width=0.49\textwidth]{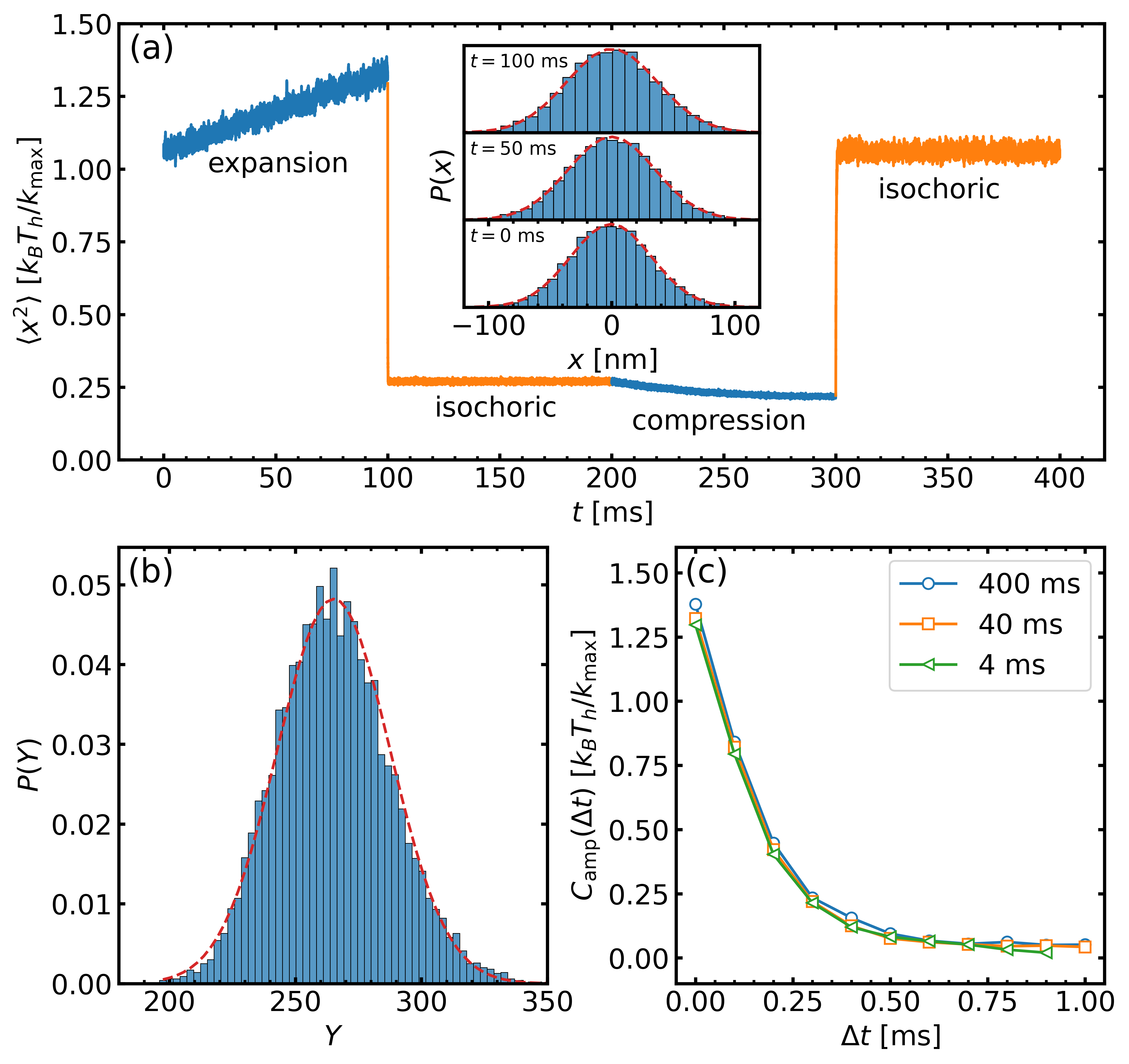}
	\caption{(Color online) (a) The position variance $\langle x^2(t) \rangle$ (ensemble average) versus time during a cycle duration time in the quasi-static regime for $\tau_{\mathrm{cyc}} = 400$ ms, and the inset shows the measured position distribution at different instants $t = 0, \ 50, \ 100$ ms. (b) The distribution of the sum of squares of random variables $Y = \sum_{i} X_{i}^2$. Here, the red dashed line is a Gaussian distribution with the same mean and variance of the distribution. (c) The amplitude of the position correlation function $C_\text{amp}(\Delta t)$ in units of $k_B T_h/k_\text{max}$ for different cycle times.}
	\label{fig:symmetry_explaination}
\end{figure}

\textit{Discussions and conclusions.}
The position of an oscillator driven by a random Brownian force is a random variable.
In the quasi-static regime, the equipartition theorem states that the stochastic position at different instants should be a Gaussian distributed random variable with the variance $\langle x^2(t)\rangle = k_{\mathrm{B}}T/k(t)$.
Figure \ref{fig:symmetry_explaination}(a) shows the position variance $\langle x^2(t) \rangle$ (ensemble average) versus time during a full cycle in the quasi-static regime, and the inset shows the measured position distribution at different instants. They agree well with the prediction from the equipartition theorem. As a result, the output work $W$ calculated via Eq. (\ref{eq: work}) would be the integration (or sum) of squares of a series of Gaussian distributed random variable $x(t)$. It is well known that, for a large number of independent random variables with an arbitrary but identical distribution, the sum of them will tend toward a Gaussian distribution. However, the distributions of $x^2(t)$ at different times are obviously not the same for the varying stiffness $k(t)$. To understand the work distribution observed in the experiments, we numerically investigated the distribution of the sum of the squares of a sequence of independent Gaussian random variables $\{X_1,\, X_2, ... \, X_{2N}\}$ with zero mean $\langle X_i \rangle = 0$ and varying variance $\langle X_i^{2}\rangle = T_i/k_i$, where $T_i$ and $k_i$ are two independent parameters respectively. 
%and calculate the sum of squares of the sequence $Y = \sum_i^{2N} X_i^2$.
An analogy to the isothermal strokes in the Stirling cycle, we set $T_i$ to be constant while $k_i$ scale linearly with $i$, i.e.,
$T_i = 1, \ k_i = \frac{(\beta - 1)i + N - \beta}{N - 1} \ \mathrm{for} \ i \leq N$ and $T_i = \alpha, \ k_i = \frac{1- \beta}{N}i + 2\beta - 1 \ \mathrm{for} \  N < i \leq 2N$,
where $\alpha = \frac{T_{\mathrm{c}}}{T_{\mathrm{h}}}$ and $\beta = \frac{k_{\mathrm{min}}}{k_{\mathrm{max}}}$ denotes the temperature ratio and stiffness ratio, respectively.
Figure \ref{fig:symmetry_explaination} (b) show the distribution of the square sum $Y = \sum_i^{2N} X_i^2$ for the same stiffness and temperature in Fig. \ref{fig:symmetry} (c), which indicates that the sum of the squares of a sequence of independent Gaussian-distributed random variables with varying variances is still a Gaussian random variable, and explains the experimental results we observed under the quasi-static condition.
As the cycle time decreases, the equipartition theorem is no longer held due to the fast-varying of stiffness. In this situation, not only is the consistency of the distribution but also the independence of variables is disrupted. We calculated the autocorrelation function $C(t,t+\Delta t) = \langle x(t)x(t + \Delta t)\rangle$ of the particle displacement during the expansion stroke. For a given $\Delta t$, the correlation function oscillates in the time domain with an amplitude $C_\text{amp}(\Delta t)$ (See SM for details), which is shown in Fig. \ref{fig:symmetry_explaination} (c) for different cycle times. One can see that the correlation time is about $1$ ms in all three cases, which is comparable to the duration of the expansion (compression) stroke in the case of rapid stiffness variation. We can thus expect that the correlation between the particle's positions at different instants will play an important role and the non-Gaussian distributed output works are therefore observed when the cycle time is set short.

In summary, we experimentally realized a nano-sized stochastic Stirling engine based on a levitated optomechanical system, where a silica particle serving as the working medium was subjected to power-varying optical tweezers and coupled periodically to the cold (hot) reservoir created by switching on (off) feedback cooling.
The experimental performance of the Stirling cycles including the work, heat, and efficiency is presented.
Our experimental results show that the output work distribution of the underdamped stochastic Stirling engine is Gaussian in the quasi-static regime, and it becomes more and more non-Gaussian as the cycle time decreases. This non-Gaussianity is qualitatively attributed to the strong correlation of the particle's position within a cycle in the non-equilibrium regime. Unfortunately, the exact probability distribution function of the output work for a fast stiffness variation in the underdamped regime is yet unknown \cite{Rademacher2022}.
%Near the non-equilibrium regime, the non-Gaussian distribution of the output work [ref] is observed, while the work distribution has a Gaussian profile in the equilibrium regime.  

% In contrast to a SHE in the overdamped environment, a non-Gaussian output work distribution is observed in our experiment. The phenomenon is confirmed as a special feature of underdamped Stirling SHE with further numerical simulations. A qualitative explanation of this non-Gaussianity is also discussed. 

The experimental study on the SHE in the underdamped regime has just begun. There are still a lot of open questions waiting to be answered. In this sense, the present work can be regarded as a preliminary exploration in this field. In the following, with an upgraded electronic controlling system, more complex investigations of the underdamped SHE such as the optimal efficiency at the max power, or dynamical behaviors of Otto cycles and Carnot cycles will be further explored. Integrating together with the ground state cooling \cite{cooling1,cooling2,cooling3} techniques demonstrated recently, our system could be directly turned into the platform for investigating real quantum mechanical nano-machines.

\textit{Acknowledgments.}
This work is supported by the Major Scientific Research Project of Zhejiang Lab (2019 MB0AD01), the Center initiated Research Project of Zhejiang Lab (Grant No. 2021MB0AL01) and Zhejiang Provincial Natural Science Foundation of China (Grant No. LQ22A040010). X.G acknowledges support from the Research Project for Young Scientists of Zhejiang Lab (No. 2020MB0AA04), Zhejiang Laboratory research Found (No. 2021MB0AL02), and the Major Project of Natural Science Foundation of Zhejiang Province (No. LD22F050002).

%\section{Supplementary Material}
%\beginsupplement

%\renewcommand{\appendixname}{Supplementary Information}

\section*{Supplemental Material}
%\appendix
\renewcommand{\thesection}{\Roman{section}}    %%%% but here
\renewcommand{\thefigure}{S\arabic{figure}}
\setcounter{figure}{0}
\renewcommand{\thetable}{S\arabic{table}}
\setcounter{table}{0}
\renewcommand{\theequation}{S\arabic{equation}}
\setcounter{equation}{0}

%\section*{Supplementary Information}

%\setcounter{equation}{0}

%\section{The first thing in SI
%	\label{sec:si1}}
%word
%\section{The second thing in SI}
%another word
%\subsection{The first sub-thing in SI}
%bla bla
%\subsection{The second sub-thing in SI}
%bla bla 2

\section{The experimental setup and protocol.}
\textit{The experimental setup.}---The experimental setup is displayed in Fig. 1 in the main text. A single-beam optical trap is created inside a vacuum chamber by a laser beam of wavelength $1064$ nm. The beam passes through an acousto-optic modulator (AOM, Gooch $\&$ Housego, 3080-199) that can adjust the optical trap power, and then is focused by a microscope objective (OBJ) of numerical aperture $\mathrm{NA} = 0.8$.
A silica particle of diameter $153.6$ nm is levitated in the center of the optical trap at a pressure of $1$ mbar and the damping rate due to collision with residual gas molecules is experimentally determined to be $\Gamma_{\mathrm{th}}/2\pi = 1.45$ kHz.
The optical potential is approximately harmonic, and the trap (mechanical) stiffness is proportional to the trap power, which can be altered by changing the radio frequency (RF) driving voltage of the AOM.
Throughout our work, we restrict our focus to the motion of the particle along the $x$ axis. By adjusting the driving voltage of the AOM, 
we vary the resulting mechanical oscillation frequency of the particle along the $x$ axis,
$\Omega/2\pi$, from $145.3$ to $160.4$ kHz.
Since the gas damping rate $\Gamma_{\mathrm{th}}$ is much smaller than the mechanical frequency, the oscillator operates in the underdamped regime. We measure the motion of the particle along the $x$ direction by detecting the scattered light using a quadrant photodetector (QPD) with a delay of approximately $200$ ns.

The motion of the particle's center of mass (COM) is influenced by the surrounding gas environment, which acts as a high-temperature reservoir (hot bath). 
To create a low-temperature reservoir (cold bath), we utilize an active feedback cooling scheme to cool the COM motion of the particle.
In this scheme, we exert a Coulomb force on the charged particle by applying a voltage to a pair of electrodes enclosing the trap.
The $x$-axis motion signal is sent through a bandpass filter (BPF, SRS, SIM965, $1$ MHz bandwidth) and a derivative circuit $\mathrm{d}/\mathrm{d}t$ (Zurich Instrument MFLI, delay $\sim 800$ ns, noise $\sim 43$ nV/$\mathrm{Hz}^{1/2}$) to provide a feedback signal proportional to velocity. This velocity-dependent feedback signal is then sent to an amplifier (AMP, noise $\sim 14$ $\mathrm{\mu}$V/$\mathrm{Hz}^{1/2}$) that modulates the voltage of the electrodes to cool the $x$-motion electrically. The noise resulting from the feedback cooling is approximately $S_{\mathrm{fd}} \approx 3\times 10^{-14}$ m/$\mathrm{Hz}^{1/2}$, which is much lower than the thermal noise $S_{\mathrm{th}} \approx 9\times 10^{-12}$ m/$\mathrm{Hz}^{1/2}$.
Using this feedback cooling, we can create a cold bath with the particle's COM temperature ranging from $30$ to $300$ K.

\begin{figure}[htbp]
	\centering
	\includegraphics[width=0.45\textwidth]{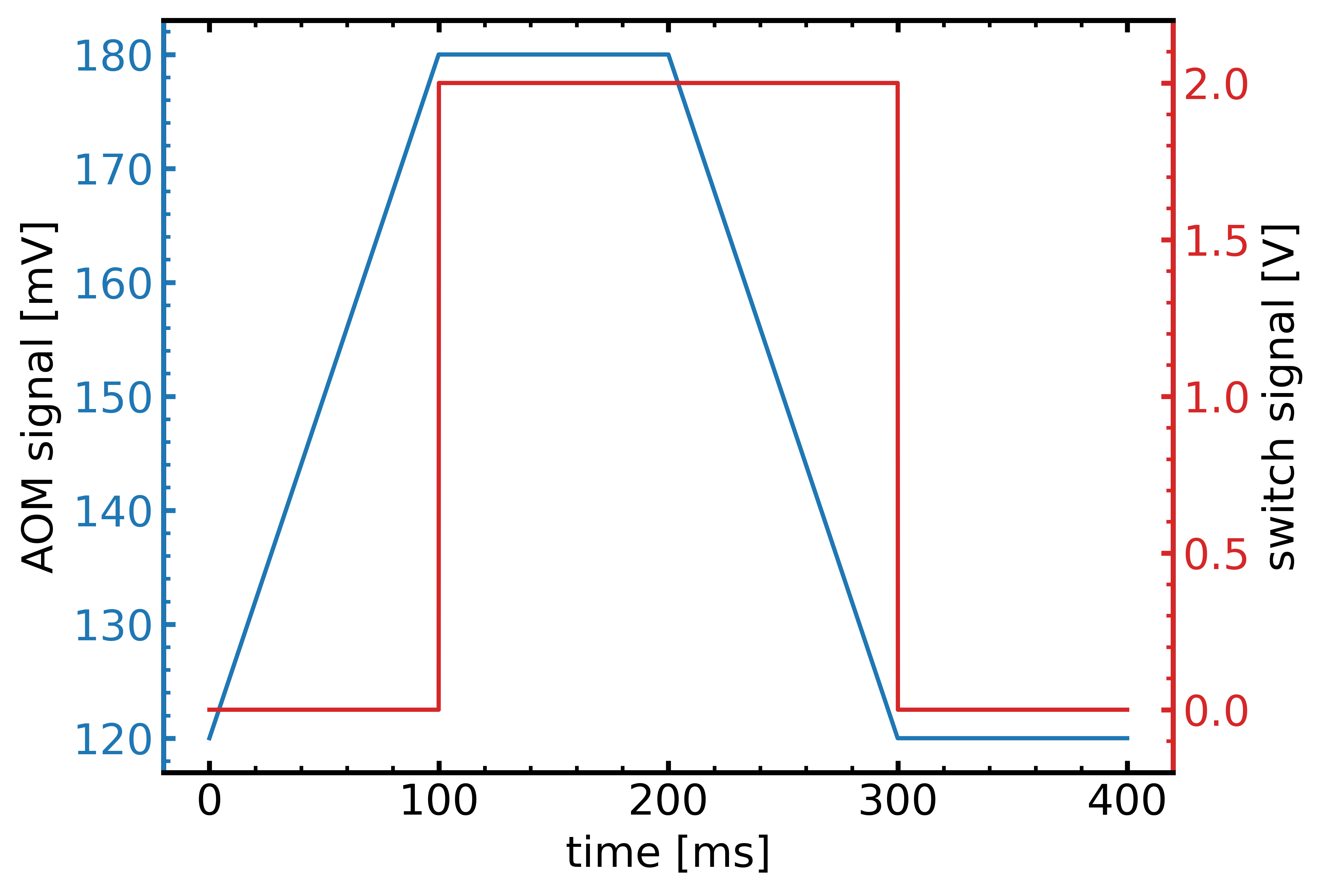}
	\caption{(Color online) The AOM and switch signals as a function of time during a Stirling cycle, where the cycle duration time is $\tau_{\mathrm{cyc}} = 400$ ms.}
	\label{fig:sg_signal}
\end{figure}

\textit{The experimental protocol of Stirling cycle.}---In order to implement the Stirling cycle, it is necessary to synchronize the changes in both the optical trap stiffness and the feedback cooling (see Fig. 1(b-c) in the main text).
To accomplish this, we employ a two-channel programmable signal generator (SG, Keysight, 33500B) that periodically outputs two synchronized signals.
One signal (AOM signal) is sent to the AOM to linearly change the optical trap power, while another signal (switch signal) is sent to a switch (delay $\sim 140 - 180$ ns) in the feedback loop to periodically turn on/off the feedback cooling. Figure \ref{fig:sg_signal} shows the AOM and switch signals as a function of time.
The driving voltage of the AOM is inversely proportional to the optical trap power, resulting in a linear decrease (increase) in the optical trap stiffness during the first (third) stroke. The switch signal is a Transistor-Transistor Logic (TTL) signal, with the feedback cooling turned on (off) when the signal is set to $2$ ($0$).

\section{The thermodynamic quantities in the quasi-static regime.}
This section presents an analytical calculation of the average thermodynamic quantities, such as work, heat, and efficiency, in the quasi-static limit. In this limit, the cycle or stroke duration time is much longer than all other relevant time scales, including the thermal relaxation time. As a result, when the protocol is changed, the system immediately adjusts to the equilibrium state corresponding to the new protocol value.

During the first isothermal stroke, the mean work done on the particle is equal to the free energy change $\Delta \mathcal{F}$ before and after the expansion, which can be expressed as
\begin{equation}
	\overline{W}_{\mathrm{h}} = \Delta \mathcal{F}_{\mathrm{h}} = \frac{k_{\mathrm{B}}T_{\mathrm{h}}}{2}\ln\frac{k_{\mathrm{min}}}{k_{\mathrm{max}}} = - \frac{k_{\mathrm{B}}T_{\mathrm{h}}}{2}\ln\frac{k_{\mathrm{max}}}{k_{\mathrm{min}}}.
\end{equation}
Similarly, the mean work done on the particle during the second isothermal stroke is given by
\begin{equation}
	\overline{W}_{\mathrm{c}} = \Delta \mathcal{F}_{\mathrm{c}} = \frac{k_{\mathrm{B}}T_{\mathrm{c}}}{2}\ln\frac{k_{\mathrm{max}}}{k_{\mathrm{min}}}.
\end{equation}
Since the two isochoric strokes in the Stirling cycle, with constant stiffness $k$, do not contribute to the work $W$, the total output is
\begin{equation}
	\overline{W} = \overline{W}_{\mathrm{h}} + \overline{W}_{\mathrm{c}} = - \frac{k_{\mathrm{B}}(T_{\mathrm{h}} - T_{\mathrm{c}})}{2}\ln\frac{k_{\mathrm{max}}}{k_{\mathrm{min}}}.
\end{equation}
We can express the output work in terms of the unit of $k_{\mathrm{B}}T_0$ as
\begin{equation}
	\overline{W} = - \frac{T_{\mathrm{h}} - T_{\mathrm{c}}}{2 T_0}\ln\frac{k_{\mathrm{max}}}{k_{\mathrm{min}}}
\end{equation}
in the main text.

To obtain the average absorbed heat in the isothermal expansion stroke, we first calculate the average internal energy change and then apply the first law.
Since the particle always remains in thermal equilibrium at a temperature $T_{\mathrm{h}}$ during this process, the average internal energy change is $\Delta U_{\mathrm{h}} = 0$. Therefore, the average absorbed heat is given by
\begin{equation}
	\overline{Q} = -\overline{W}_{\mathrm{h}} = \frac{k_{\mathrm{B}}T_{\mathrm{h}}}{2}\ln\frac{k_{\mathrm{max}}}{k_{\mathrm{min}}}.
\end{equation}
Using the total output work and the absorbed heat, we can calculate the mean efficiency of the Stirling cycle as:
\begin{equation}
	\eta = - \frac{\overline{W}}{\overline{Q}} = 1 - \frac{T_{\mathrm{c}}}{T_{\mathrm{h}}} = \eta_{\mathrm{Carnot}},
\end{equation}
where $\eta_\mathrm{Carnot}$ is the efficiency of a Carnot cycle operating between the same two temperatures $T_\mathrm{h}$ and $T_\mathrm{c}$.

\section{Numerical simulation.}
\textit{The dimensionless Langevin equation.}---We introduced dimensionless quantities $\Tilde{\Omega} \equiv \Omega/\omega$, $\Tilde{t} \equiv \omega t$, $\Tilde{\Gamma}_{\mathrm{th/fd}} \equiv \Gamma_{\mathrm{th/fd}}/\omega$, $\Tilde{k} \equiv \Tilde{\Omega}^2 = k/\omega^2$, and $\Tilde{x} = x/\sqrt{\frac{k_{\mathrm{b}}T}{m\omega^2}}$, and obtain the following dimensionless Langevin equation,
\begin{equation}\label{eq:dimensionlessLEQ}
	\ddot{\Tilde{x}} + (\Tilde{\Gamma}_{\mathrm{th}} + \Tilde{\Gamma}_{\mathrm{fd}}) \dot{\Tilde{x}} + \Tilde{k} \Tilde{x} = \sqrt{2 \Tilde{\Gamma}_{\mathrm{th}}} \xi(\Tilde{t}).
\end{equation}
Here, we set the initial maximum mechanical frequency to be the normalized unit, i.e., $\omega = \Omega_{\mathrm{max}}$.
According to the experimental parameters in the main text, the related dimensionless parameters for the simulation are given by
$\Tilde{\Omega}_{\mathrm{min}} = \frac{145.3}{160.4} \simeq 0.906$, $\Tilde{\Gamma}_{\mathrm{h}} = \Tilde{\Gamma}_{\mathrm{th}} = \frac{1.45}{160.4} \simeq 0.009$, and $\tilde{\tau}_{\mathrm{cyc}} = \Omega_{\mathrm{max}} \tau_{\mathrm{cyc}}$.
The damping rate of the cold bath is obtained via $\Tilde{\Gamma}_{\mathrm{c}} = \Tilde{\Gamma}_{\mathrm{h}}/\alpha$ with the temperature ratio $\alpha =  T_{\mathrm{c}}/T_{\mathrm{h}}$ and $T_{\mathrm{h}} = 320$ K, for instant, $\Tilde{\Gamma}_{\mathrm{c}} \simeq 0.044$ for the cold bath temperature $T_{\mathrm{c}} = 65$ K.

\begin{figure}[htbp]
	\centering
	\includegraphics[width=0.45\textwidth]{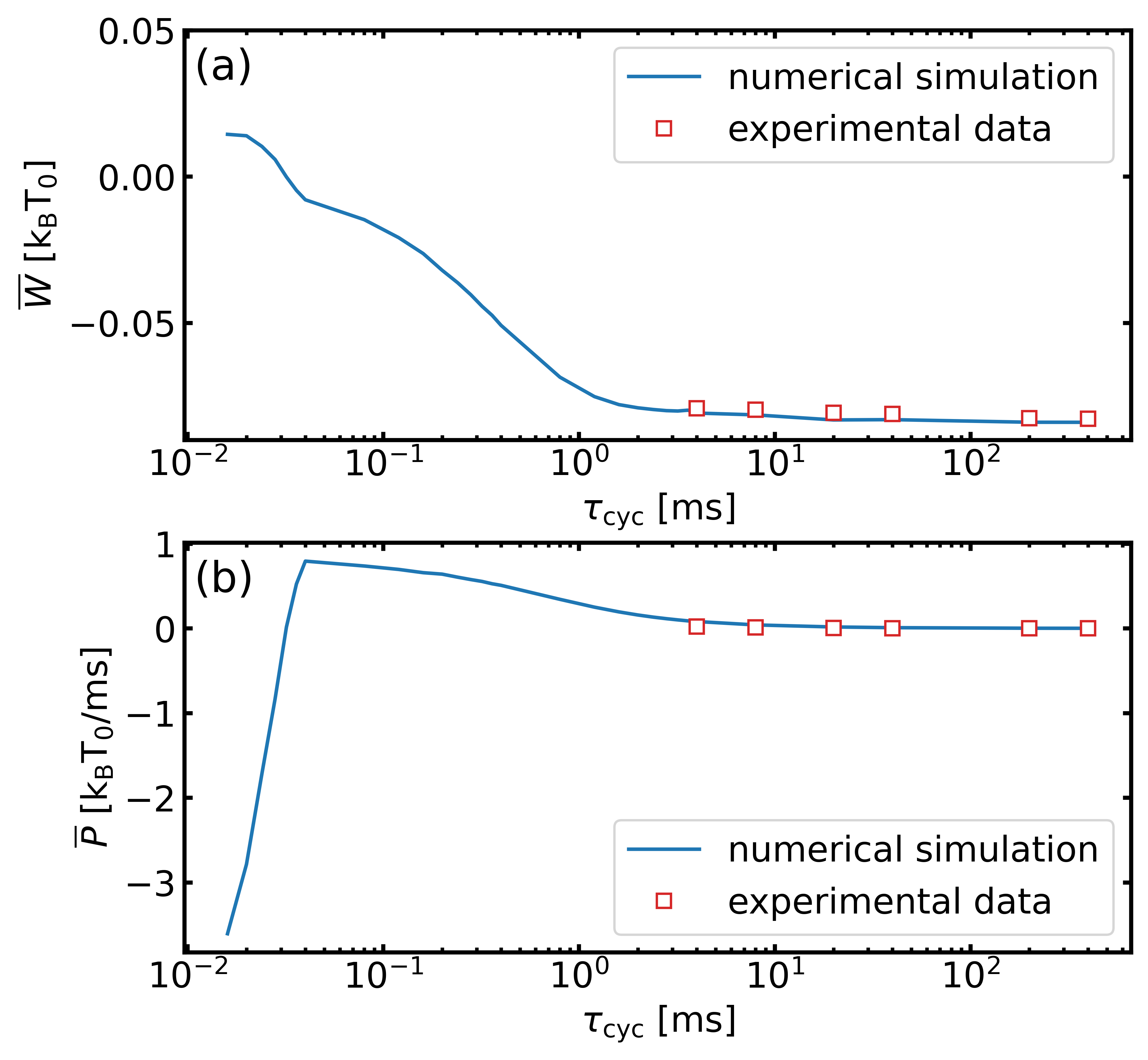}
	\caption{(a) The mean output work and (b) the mean power as a function of the cycle duration time $\tau_{\mathrm{cyc}}$.}
	\label{fig:power}
\end{figure}

\textit{The maximum power.}---We numerically simulated the Stirling cycle via the dimensionless Langevin equation (\ref{eq:dimensionlessLEQ}), where the cold bath temperature is set to $T_{\mathrm{c}} = 65$ K. Figure \ref{fig:power} shows the mean output work and the mean power as a function of the cycle time. 
When the cycle time is very small, the mean output work is positive indicating the work done on the environment. As the cycle time increases, the mean output work also increases and gradually approaches the theoretical value of the quasi-static regime.
The mean power initially increases to a maximum value, then begins to decrease, and eventually approaches $0$. The maximum power is achieved when the cycle time $\tau_{\mathrm{cyc}} = 0.04$ ms.
The experimental results are in good agreement with the numerical simulation.

\section{Jensen–Shannon divergence.}

Jensen–Shannon divergence (JSD) \cite{JSD2, JSD3}, a symmetrized and smoothed version of the Kullback–Leibler divergence (KLD), is a quantity of measuring the similarity between two probability distributions. For two probability distribution $P = (p_1, p_2, ..., p_n)$ and $Q = (q_1, q_2, ..., q_n)$ with $\sum_{i}^{n}p_i = \sum_{i}^{n}q_i = 1$, the JSD is defined as
\begin{equation}
	D_{\mathrm{JS}}(P, Q) = \frac{D_{\mathrm{KL}}(P, M) + D_{\mathrm{KL}}(Q, M)}{2},
\end{equation}
where $M = (P + Q)/2$ and
\begin{equation}
	D_{\mathrm{KL}}(P, Q) = \sum_{i}^{n}p_{i}\log\left(p_i/q_i\right)
\end{equation}
is the Kullback–Leibler divergence.
The Jensen–Shannon divergence is bounded in $0 \leq D_{\mathrm{JS}}(P, Q) \leq 1$ and only vanishes when $P = Q$.

\section{The autocorrelation functions.}

The autocorrelation function (ACF) is a statistical tool that quantifies the degree of correlation between a variable and its past values. In our experiments, we use the ACF of the particle displacement $x(t)$, which is calculated as 
\begin{equation}
	C(t, t + \Delta t) = \langle x(t)x(t + \Delta t) \rangle,
\end{equation}
where $\Delta t$ represents the time lag and $\langle \dots \rangle$ denotes  the average over all realizations.

\begin{figure}[htbp]
	\centering
	\includegraphics[width=0.45\textwidth]{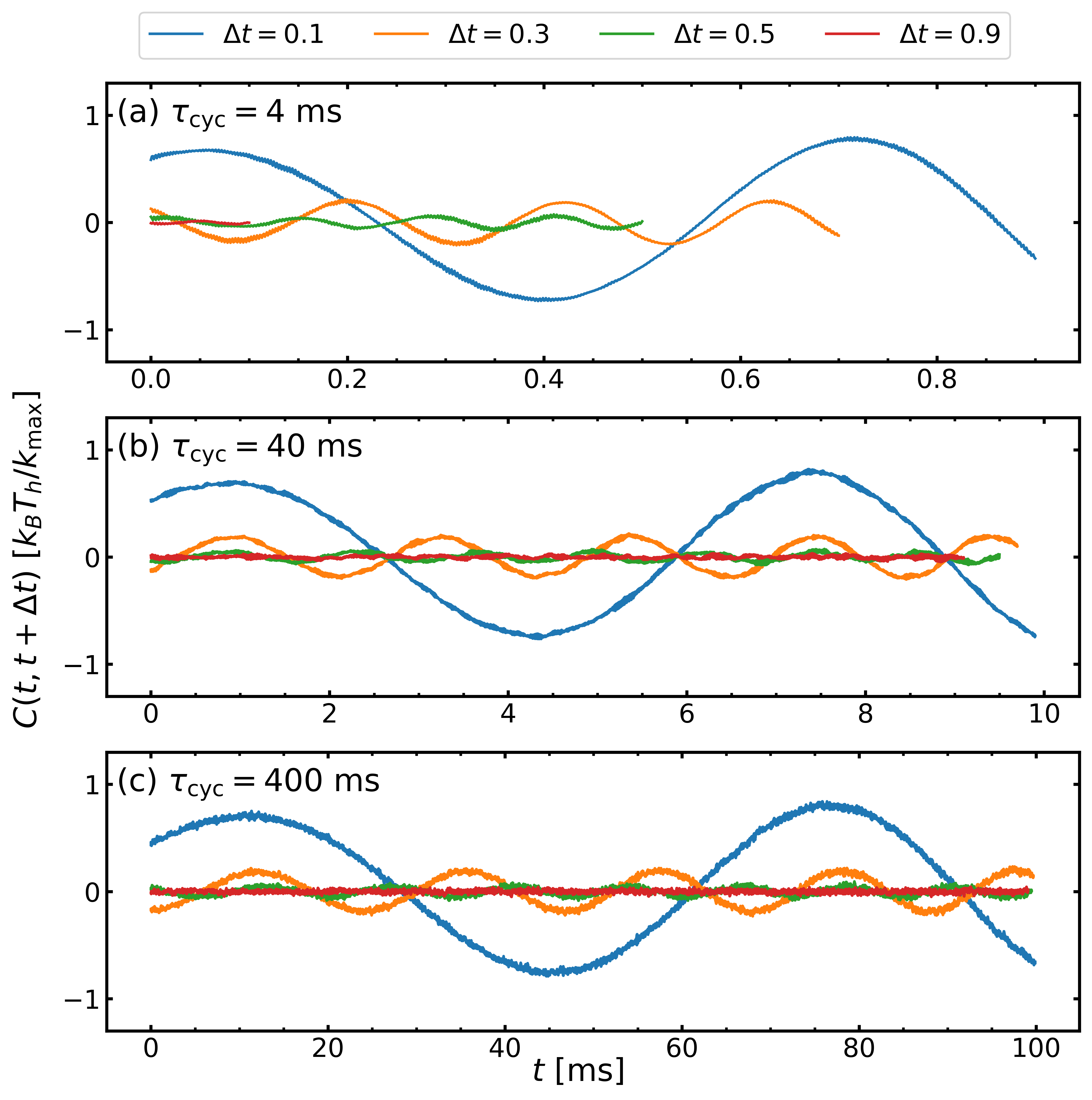}
	\caption{The ACFs of the experiment data as a function of time $t$ for different time lags and cycle time during the isothermal expansion stroke. The other parameters are the same as that in Fig. 3(a-c) in the main text.}
	\label{fig:correlation}
\end{figure}

Figure \ref{fig:correlation} displays the ACFs of the experiment data as a function of time $t$ for different time lags and cycle time during the isothermal expansion stroke. It shows that the ACFs exhibit an oscillation behavior with time $t$. For all the cycle time (i.e., $\tau_{\mathrm{cyc}} = 4, 40, 400$ ms), the amplitudes of the ACFs decrease as the time lag $\tau$ increases, and they approach zeros for a time lag of $\tau = 0.9$ ms. 

To more clearly decrease the dependence between the ACF and the time lag, we introduce the amplitude of the ACFs defined as follows:
\begin{equation}
	C_\text{amp}(\Delta t) = \max\{|C(t, t + \Delta t)|\}.
\end{equation}
This new definition of the ACF focuses solely on the time lag, enabling us to analyze better how correlations change with the cycle time and time lag.

%To more clearly decrease the dependence between the ACF and the time lag, we introduce the amplitude of the ACFs defined as follows:
%\begin{equation}
%	C_\text{amp}(\Delta t) = \max\{|C(t, t + \Delta t)|\}.
%\end{equation}
%This new definition of the ACF focuses solely on the time lag, enabling us to analyze better how correlations change with the cycle time and time lag.

\bibliography{ref_prl.bib}
\end{document}